# Energetic Particle Instabilities in Fusion Plasmas

S.E. Sharapov[1], B. Alper[1], H.L. Berk[2], D.N. Borba[3], B.N. Breizman[2], C.D. Challis[1], I.G.J. Classen[4], E.M. Edlund[5], J. Eriksson[6], A. Fasoli[7], E.D. Fredrickson[5], G.Y. Fu[5], M. Garcia-Munoz[8], T. Gassner[9], K. Ghantous[5], V. Goloborodko[9], N.N. Gorelenkov[5], M.P. Gryaznevich[1], S. Hacquin[10], W.W. Heidbrink[11], C. Hellesen[6], V.G. Kiptily[1], G.J. Kramer[5], P. Lauber[8], M.K. Lilley[12], M. Lisak[13], F. Nabais[3], R. Nazikian[5], R. Nyqvist[13], M. Osakabe[14], C. Perez von Thun[8,15], S.D. Pinches[16], M. Podesta[5], M. Porkolab[17], K. Shinohara[18], K. Schoepf[9], Y. Todo[14], K. Toi[14], M.A. Van Zeeland[19], I. Voitsekhovich[1], R.B. White[5], V. Yavorskij[9], ITPA EP TG, and JET-EFDA Contributors[*]

[1]Euratom/CCFE Fusion Association, Culham Science Centre, Abingdon, OX14 3DB, UK
[2]Institute for Fusion Studies, University of Texas at Austin, Austin, Texas 78712, USA
[3]Assoc. EURATOM/IST, Inst. de Plasmas e Fusão Nuclear - IST, UTL, P-1049-001 Lisboa, Portugal
[4]FOM institute DIFFER – Assoc. Euratom-FOM, Nieuwegein, The Netherlands
[5]Princeton Plasma Physics Laboratory, Princeton NJ 08543, USA
[6]Dept. Physics & Astronomy, Uppsala University, Euratom/VR Fusion Assoc., Uppsala, Sweden
[7]CRPP/EPFL, Association Euratom-Confederation Suisse, Lausanne, Switzerland
[8]Max Planck Institute für Plasmaphys., Euratom Assoc., Boltsmannstr.2, D-85748 Garching, Germany
[9]Euratom/OEAW Fusion Assoc., Inst. Theoretical Physics, Univ. of Innsbruck, A-6020, Austria
[10]CEA, IRFM, Assoc. Euratom-CEA, St Paul Lez Durance, France
[11]Dept. of Physics and Astronomy, University of California, Irvine, CA 92697, USA
[12]Physics Department, Imperial College, London, SW7 2AZ, UK
[13]Dep. of Earth and Space Science, Chalmers Univ. of Technology, SE-412 96 Goteborg, Sweden
[14]National Institute for Fusion Science, 322-6 Oroshi-cho, Toki 509-5292, Japan
[15]JET-EFDA Close Support Unit, Culham Science Centre, Abingdon, OX14 3DB, UK
[16]ITER Organization, Route de Vinon sur Verdon, 13115 St Paul-lez-Durance, France
[17]Plasma Science and Fusion Centre, MIT, Cambridge, Massachusetts 02139, USA
[18]Japan Atomic Energy Agency, Naka, Ibaraki 319-0193, Japan
[19]General Atomics, PO Box 85608 San Diego, CA 92186-5608, USA.

*See the Appendix of F. Romanelli et al., Proc. 24th IAEA Fusion Energy Conference 2012, San Diego, USA

E-mail: Sergei.Sharapov@ccfe.ac.uk

**Abstract.**
Remarkable progress has been made in diagnosing energetic particle instabilities on present-day machines and in establishing a theoretical framework for describing them. This overview describes the much improved diagnostics of Alfvén instabilities and modelling tools developed world-wide, and discusses progress in interpreting the observed phenomena. A multi-machine comparison is presented giving information on the performance of both diagnostics and modelling tools for different plasma conditions outlining expectations for ITER based on our present knowledge.

# 1. Introduction

As energetic alpha particles will play a central role in burning deuterium-tritium (DT) plasmas, it is crucial to understand and possibly control their behaviour in various operational regimes. Of particular importance is the understanding of instabilities driven by alpha particles [1]. The complete set of implications for operating burning plasmas with the alpha-particle driven instabilities can only be investigated in a burning plasma experiment itself. However, experiments on present day machines with energetic particles produced by neutral beam injection (NBI), ion-cyclotron resonance heating (ICRH), and electron-cyclotron resonance heating (ECRH) already reveal many relevant features of the possible alpha-particle instabilities. The energetic particle-driven instabilities are often observed experimentally and they range from low-frequency fishbones in the range of 10-50 kHz up to compressional Alfvén eigenmodes (CAEs) in the frequency range comparable to or higher than the ion cyclotron frequency. The instability of weakly-damped Alfvén eigenmodes (AEs) is of highest priority for the next step burning plasma on ITER due to a number of reasons. First, AEs are driven by radial gradient of energetic particle pressure and lead to enhanced alpha-particle radial transport, in contrast to CAEs excited by velocity space gradients. Second, AEs resonate with alpha-particles in the MeV energy range in contrast to, e.g. fishbones expected to resonate with alpha-particles of 300-400 keV in ITER. Third, due to their weak damping, AEs could be excited by alpha-particle population with lower energy content per volume as compared to linear Energetic Particle Modes (EPMs). Although amplitudes of Alfvén perturbations are usually not too high in present day experiments, the existing experimental data on energetic particle radial redistribution and losses is representative enough to gain important information on the processes involved.

A significant dedicated effort was made in the past decade in developing techniques of diagnosing energetic particle-driven Alfvén instabilities with interferometry, ECE, phase contrast imaging, and beam emission spectroscopy. Together with the much improved diagnostics of the energetic ions themselves, such development sets a new stage for understanding of such instabilities since nearly all the essential information can now be obtained from experimental measurements and not from assumptions or modelling with often uncertain error bars. The aim of this overview is to present a point-by-point comparison between the much improved diagnostics of AEs and modelling tools developed world-wide, and outlines progress in interpreting the observed phenomena.

Experimentally, Alfvén instabilities exhibit two main nonlinear scenarios, with a mode

Frequency Locked (FL) to plasma equilibrium, or with a mode Frequency Sweeping (FS) ("frequency chirping" modes [2]). It is important to understand these two scenarios for predicting what temporal evolution and transport due to Alfvén instabilities will be relevant to ITER. Figure 1 presents a typical example of FL Alfvén eigenmodes (AEs) on the JET tokamak with ICRH-accelerated ions [3], while Figure 2 presents FS Alfvén instability on the JT-60U tokamak with negative NBI heating [4]. In the case of JET, the Alfvén perturbations form a discrete spectrum of Toroidal Alfvén Eigenmodes (TAEs) with different toroidal mode numbers $n$ and frequencies, which are determined by bulk plasma equilibrium throughout the whole nonlinear evolution. These TAEs with different $n$'s appear one-by-one as ICRH power increases, and the observed slow change in TAE frequencies is caused by an increase in plasma density in accordance with Alfvén scaling $V_A \propto B/\sqrt{n_i(t)}$. The amplitude of each TAE saturates and remains nearly constant. In contrast to the FL scenario, Figure 2 shows FS Alfvén instability on JT-60U with frequency of the perturbations starting from TAE frequency, but then changing on a time scale much shorter than the time scale of the changes in plasma equilibrium. The amplitude of this FS instability exhibit bursts, and the mode frequency sweeps during every burst.

The FL and FS scenarios of energetic particle-driven instabilities differ in the temporal evolution of redistribution and losses of energetic particles and they require conceptually different approaches in modelling. Namely, the frequency of unstable modes in FL scenario correspond to linear AE determined by bulk plasma equilibrium throughout the linear exponential growth and nonlinear evolution of the mode. In this case, the energetic particles determine growth rate, but affect very little the eigenmode structure and frequency, so that the modes are "perturbative". In FS scenario, the contribution of the energetic particles to mode frequency is as essential as the bulk plasma contribution, and when the unstable mode re-distributes the energetic particles, it changes the frequency too. Characteristic time scale of the energetic particle redistribution is the inverse growth rate, so the energetic particle profile and the mode frequency determined by this profile change much faster than the plasma equilibrium. The energetic particles cannot be considered as a small perturbation in the FS scenario, so the modes are "non-perturbative" nonlinear energetic particle modes.

In the past studies of FL scenarios linear spectral MHD codes could be used for computing AEs supported by the plasma equilibrium. For FL scenarios observed experimentally, MHD spectroscopy via AEs, i.e. obtaining information on plasma equilibrium from observed spectrum of AEs, became possible [5-7]. In particular, Alfvén cascade (AC) eigenmodes [6,8] (also called reversed shear Alfvén eigenmodes, RSAEs [9]) were employed in MHD spectroscopy successfully. In contrast to TAE modes in Fig.1, ACs exist in plasmas with reversed magnetic shear in the early phase of discharges, when the plasma current is not fully penetrated and the safety factor $q(t) \approx rB_T/(RB_P(t))$ evolves in time. In this case, ACs are localised at the minimum of the safety factor, $q_{min}$, and frequency of AC locks to the time dependent $q_{min}(t)$ as $\omega_{AC} \approx |n - m/q_{min}(t)| \cdot V_A/R + \Delta\omega$ [8].

For the FS scenario, the concept of near-threshold "hard" nonlinear regime of energetic particle-driven instability has demonstrated the possibility of forming non-perturbative nonlinear modes even when the instability is somewhat below the linear threshold. This recent development began to provide a credible opportunity of understanding FS modes to a degree required for theory-to-experiment comparison and predictions for burning plasmas.

## 2. Advances in diagnosing Alfvén instabilities

Recent advances in diagnosing Alfvén instabilities are associated with a significant expansion of tools and techniques for detection and identification of the unstable modes. In the past, Alfvén instabilities were detected via perturbed magnetic field measured by magnetic sensors, e.g. Mirnov coils outside the plasma. Such measurements did not always detect AEs in the plasma core and they will be more difficult in ITER and DEMO due to the necessity of protecting the magnetic sensors. It is also desirable for future DT machines with a restricted access to the plasma to have detection systems for Alfvén instabilities naturally combined with some other diagnostic tools. Measurements of perturbed electron density and temperature associated with AEs are possible alternatives to magnetic sensors at the edge. The perturbed electron density caused by AEs in toroidal geometry is given by

$$\frac{\delta n}{n_0} = -\xi \cdot \frac{\nabla n_0}{n_0} - \nabla \cdot \xi = \left(\frac{\hat{\mathbf{n}}}{L_n} - 2\frac{\hat{\mathbf{R}}}{R^2}\right) \cdot \xi , \qquad (1)$$

where $\delta n$, $n_0$ are the perturbed and equilibrium densities, $\xi$ is the plasma displacement, and $L_n$ is the radial scale length of the density. The first term in Eq.(1) describes the usual convection of plasma involved in the $\mathbf{E} \times \mathbf{B}$ drift. The second term $\propto 1/R$ in (1) is caused by toroidicity and gives a non-zero perturbed density when the profile of $n_0$ is flat [10, 11].

A launched microwave O-mode beam on JET with frequency above the cut-off frequency of O-mode was found to deliver detection of AEs far superior to that made with Mirnov coils [11]. This "O-mode interferometry" shows unstable AEs not seen with Mirnov coils. Later, the standard far infra-red (FIR) JET interferometer was digitised to high sampling rate, which enabled detecting AEs even in plasmas of high density. A similar interferometry technique was employed for diagnosing AEs in DIII-D discharges [12]. It was observed for the first time that a "sea of modes" exists in such plasmas with toroidal mode numbers up to $n = 40$.

The interferometry technique has increased significantly the sensitivity of AE detection and it assures that all unstable modes are detected even deeply in the plasma core. Since the interferometry technique of detecting AEs requires only interferometers used for plasma density measurements, this method is a good candidate for ITER and DEMO.

The main limitation of using interferometry or Mirnov coils for detecting AEs, is that the AEs cannot be localised from the measurements. Recent successful development of ECE [13] and ECE imaging [14,15], beam emission spectroscopy (BES) [16], and phase contrast imaging (PCI) [17] have addressed the problem of measuring mode structure. Together with the existing SXR technique and X-mode reflectometry used for observing alpha-driven AEs in DT plasmas [10], the new diagnostics provide opportunities in identifying the spatial structure of the modes to a degree required for an accurate experiment-to-theory comparison.

On ALCATOR C-MOD, the PCI diagnostic was found to be an outstanding tool for detecting core-localised AEs [17]. This diagnostic is a type of internal beam interferometer, which can generate a 1D image decomposed in 32 elements of approximately 4.5 mm chord separation in the direction of major radius thus providing information on AE localisation.

On DIII-D and ASDEX-Upgrade, ECE became a successful tool for measuring AEs. Figures 3, 4 display an example of the ECE radial profiles for beam-driven ACs (RSAEs) and TAEs in DIII-D discharge [13]. Both localisation and the radial widths of these FL modes are found to agree well with the linear MHD code NOVA, which also includes the relationships between the perturbed magnetic fields, density and electron temperature. The perturbed electron temperature associated with the modes is estimated to be $\delta T_e / T_e \approx 0.5\%$, while the perturbed electron density from BES is found to be $\delta n_e / n_e \approx 0.25\%$. By comparing the measurements shown in Fig. 4 with NOVA calculations, one deduces that the peak values of the perturbed magnetic field are $\delta B / B \approx (1.5 \pm 0.14)\ 10^{-4}$ for RSAE and $\delta B / B \approx (2.5 \pm 0.18)\ 10^{-4}$ for TAE. This information is necessary for computing the energetic particle redistribution due to the AEs described in the next Section.

In addition to the Alfvén diagnostics, there has been an extensive development in diagnostics of confined and lost energetic particles on many machines world-wide [18]. Description of these diagnostics goes beyond the scope of this paper, but some examples of their use will be presented.

## 3. Redistribution and losses of energetic ions caused by Alfvén instabilities

It was noted in the previous Section that typical amplitudes of the AEs excited are quite low, e.g. in the range of $\delta B / B \approx 10^{-4} \div 10^{-3}$ on the DIII-D tokamak. For such amplitudes, particles could be affected noticeably if motion of such particles is in resonance with the wave. Hence, the relatively narrow regions surrounding the wave-particle resonances are of major importance for describing the particle interaction with AEs. Significant effort has been made in order to validate experimentally the main assumptions and results of both linear and nonlinear theory describing resonant interaction between Alfvén waves and energetic particles, and the effect of Alfvén instabilities on redistribution and losses of the energetic particles. In a tokamak, the theory focuses on the dynamics of particles resonant with a wave, i.e. satisfying the resonance condition (in the guiding centre approximation)

$$\Omega \equiv \omega - n\omega_\varphi(E,\ P_\varphi,\ \mu) - l\omega_\vartheta(E,\ P_\varphi,\ \mu) = 0,\ l = 0,\ \pm 1,... \tag{2}$$

Here, the toroidal, $\omega_\varphi(E,\ P_\varphi,\ \mu)$, and poloidal, $\omega_\vartheta(E,\ P_\varphi,\ \mu)$, orbit frequencies of the particles in the unperturbed fields are functions of three invariants: energy $E$, magnetic moment $\mu$, and toroidal angular momentum,

$$P_\varphi \equiv -(e_\alpha/2\pi c)\Psi(r) + m_\alpha R V_\parallel B_\varphi/B, \tag{3}$$

where $\Psi(r)$ is the poloidal flux, $V_\parallel$ is the velocity of the particle parallel to the magnetic field, $e_\alpha$, $m_\alpha$ are the charge and mass of the particle, $B_\varphi$ is the toroidal component of the magnetic field, and $R$ is the major radius. Since the wave frequency is much less than ion cyclotron frequency, $\mu$ is conserved as is the combination $E - (\omega/n)P_\varphi$ (for a single mode) giving

$$\Delta E = (\omega/n)\Delta P_\varphi. \tag{4}$$

The free energy source of the Alfvén instability is associated with radial gradient of energetic particle pressure and causes the wave growth rate

$$\gamma_L/\omega \propto q^2 r(d\beta_h/dr)\cdot F(V_A/V_h)\cdot G(\Delta_{AE}/\Delta_h); \tag{5}$$

where $\beta_h$, $V_h$, $\Delta_h$ are the beta value, thermal velocity, and the drift orbit width of the energetic (hot) particles, $\Delta_{AE}$ is the radial width of the mode, and functions $F(V_A/V_h)$ and $G(\Delta_{AE}/\Delta_h)$ depend on the energy distribution of the energetic particles.

The energetic particle-driven Alfvén instability develops if the growth rate (5) exceeds the absolute value of the damping rate of the wave, $\gamma_d$, to give a positive net growth rate

$$\gamma \equiv \gamma_L - \gamma_d > 0. \tag{6}$$

Existence of AEs with very low values of $\gamma_d$ was validated in the studies of plasma response to externally launched waves with frequencies scanned over the Alfvén range on JET [19] and C-MOD [20]. High-quality resonances associated with TAE and EAE, with $Q \equiv \omega/\gamma_d \approx 10^2 - 10^3$, were found. Recently, dedicated new TAE antennae were installed on JET [21] and MAST [22] for studying stable AEs with low $\gamma_d$ over a wider parameter range, and ITPA benchmark/validation studies on the values of $\gamma_d$ were performed for ITER [23].

For assessing transport caused by AEs, one notes that each mode affects resonant particles only in a relatively narrow region of the phase space indicated by condition (2), and that AE can cause a significant radial redistribution of these particles with a minor change in their energy (see Eq.(4)). In the nonlinear phase of instability, the resonant particles can become trapped in the field of the wave within a finite width of the resonance, $\Delta\Omega \cong \omega_{NL}$, where $\omega_{NL}$ is the nonlinear trapping frequency [24]. The nonlinear width of the resonance varies along the resonant surface depending on the unperturbed particle orbits, the mode structure, and the mode amplitude. If the widths of different resonances are smaller than the distance between them, a single mode nonlinear theory applies. If the resonances overlap, stochastic diffusion of the particles over many resonances can cause a global transport [24, 25].

Two representative cases of AE-induced redistribution of the energetic particles, with resonances non-overlapped and overlapped, were recently modelled in detail for well-diagnosed experiments on JET and DIII-D. In both cases FL scenarios are relevant, so the structure and frequencies of the AEs could be obtained from a linear theory.

On JET, D beam ions were accelerated from 110 keV up to the MeV energy range by 3$^{rd}$ harmonic ICRH in D plasmas [26]. Figure 5 displays the temporal evolution of the main plasma parameters. The yield of DD fusion reaction increases by a factor of ~6 when ICRH is applied (12−14 s) indicating that a significant fraction of the beam ions were accelerated to energies in the MeV energy-range where the DD reactivity is almost two orders of magnitude higher than at the beam injection energy. This was confirmed by neutron spectroscopy measurements by TOFOR [27] where a spectrum of DD neutrons reaching up to ~6 MeV was measured with ~ 250 msec time resolution. Such high neutron energies from the DD reaction in turn requires D ions with energies of up to 3 MeV. A monster sawtooth is formed at $\approx 14$ sec, with both Mirnov coils and the FIR interferometry detecting "tornado" modes (TAE inside the $q = 1$ radius [28]) soon after as Figure 6 shows. The frequency evolution of the tornadoes is determined by the decrease of $q(0)$ before the monster sawtooth crash at $\approx 15.6$ s [28]. The modes correlate with a significant radial redistribution of ICRH-accelerated D ions in the plasma core. The diagnostic showing the profile of D ions with energy $E > 0.5$ MeV, which includes the phase space region satisfying the resonance condition (2), is based on γ-rays from the nuclear $^{12}C(D,p)^{13}C$ reaction between C impurity and fast D [29]. During the tornado modes, the 2D γ-camera on JET (Fig.7) measuring the γ-emission with time resolution of ~ 50 msec showed a strong redistribution of the γ-emission in the plasma core as Figure 8 displays.

A suite of equilibrium (EFIT and HELENA) and spectral code MISHKA was used to model the observed AEs. The particle-following code HAGIS [30] was then employed to simulate the interaction between the energetic ions and TAEs. The unperturbed distribution function of fast D ions was assumed to be of the form $f(E, P_\varphi, \mu) = f(E) \cdot f(P_\varphi) \cdot f(\Lambda)$, where the distribution function in $\Lambda \equiv \mu B_0 / E$ for trapped energetic ions accelerated with on-axis ICRH was considered to be Gaussian centred on $\Lambda = 1$ with the width of $\Delta \Lambda \approx 1.5 \cdot 10^{-1}$. The distribution function in energy, $f(E)$, was derived from the measured energy spectrum of DD neutrons [27]. The spatial profile of the trapped D ions before the TAE activity, $f(P_\varphi)$, was obtained with the best fit matching the observed 2D profile of the gamma-emission.

The initial value simulation with HAGIS shows an exponential growth of the modes followed by nonlinear saturation and redistribution of the trapped energetic ions. Figure 9 demonstrates that these HAGIS results are in satisfactory agreement with the experimentally measured gamma-ray profiles for the trapped energetic ions. In view of the possible interplay between TAE and sawteeth [31], the fast particle contribution to the stabilising effect of the $n=1$ kink mode was computed before and after the redistribution. Significant decrease in the stabilising effect was found in [26] supporting the idea that monster crash is facilitated by TAEs expelling the energetic ions from the region inside the $q=1$ radius. A similar interplay between TAE modes redistributing energetic particles in the plasma core, and other types of MHD instabilities affected by the energetic particles, could be relevant for ITER. Since the interaction described above does not require energetic particle transport across the whole radius of the plasma, even transport of the energetic particles deeply in the plasma core could affect MHD stability in the same core region.

Another example of energetic particle redistribution by AEs comes from DIII-D experiments showing significant modification of D beam profile in the presence of multiple TAEs and RSAEs (Alfvén cascades) [32]. This observation and TRANSP predictions are shown in Figure 10, and TAE and RSAE data is shown in Figs 3, 4. Based on the ECE measurements of AE amplitude and mode structure, accurate modelling was performed in [33] with the ORBIT code [25] to interpret the flat fast ion profile. In this case of multiple modes densely packed in frequency, a wide area of wave-particle resonance overlaps was found. A stochastic threshold for the beam transport was estimated, and the experimental amplitudes were found to be only slightly above this threshold [33]. We thus observe that multiple low amplitude AEs can indeed be responsible for substantial central flattening of the beam distribution as Figure 11 shows.

Recently, a quasi-linear 1.5D model has been developed and applied to this DIII-D data [34]. The model gives the relaxed fast ion profile determined by the competition between the AE drive and damping. Figure 12 presents a comparison between the experimental data, the TRANSP modelling, and the 1.5D quasi-linear model [34] for the slowing-down distribution function of the beam. It is important to point out that the ratio between the beam orbits and minor radius in the DIII-D experiments, $\Delta_h/a \cong 0.1$, was well above the value expected for ITER, $\Delta_h/a \cong 10^{-2}$. This difference implies that a similar excitation of RSAEs localised at $q_{\min}$ in ITER will not necessarily give a global transport similar to that observed on DIII-D. In another theory-to-experiment comparison aimed at explaining TFTR results [35], the role of nonlinear sidebands including zonal flows was shown to be significant in reducing the mode saturation level. The beam ion losses caused by the AEs were proportional to $(\delta B)^2$, although the simulation cannot yet match the experimental data precisely.

In ASDEX-Upgrade experiments, detailed measurements of radial structures of AE driven by beams and ICRH were obtained using ECE imaging [14], SXR, and reflectometry. The fast-ion redistribution and loss is routinely monitored with scintillator based fast-ion loss detectors and fast-ion D-alpha spectroscopy. It was found that a radial chain of overlapping AEs enables the transport of fast-ions from the core all the way to the loss detector [36, 37].

## 4. MHD spectroscopy of plasma

The FL instabilities of AEs represent an attractive form of MHD spectroscopy [5-7] since AEs are usually numerous, easily detectable, and do not cause degradation of plasma or the fast ion confinement as long as the AE amplitudes are sufficiently small. Last, but not least, AEs in ITER may provide a natural passive diagnostic tool of the burning plasma equilibrium. The most impressive result of MHD spectroscopy based on AEs was associated with AC eigenmodes [6,8,38] excited in reversed-shear plasmas prepared for triggering internal transport barriers (ITBs). In discharges with low auxiliary power per volume as in JET, diagnosis of the time evolution of $q_{\min}(t)$ was important for developing reproducible scenarios with ITBs [39]. MHD spectroscopy was found to be the best option for that, since AC eigenfrequency, $\omega_{AC}$, traces the evolution of $q_{\min}(t)$ as:

$$\frac{d}{dt}\omega_{AC}(t) \approx m\frac{V_A}{R}\frac{d}{dt}q_{\min}^{-1}(t). \qquad (7)$$

Here, $m$ is poloidal mode number of an AC and $V_A$ is Alfvén velocity.

In addition to the scenario development, important information was obtained on the time sequence of events causing the ITB. Figure 13 shows a typical JET discharge, in which a grand AC with all mode numbers $n$ seen at once signifies that $q_{\min}$ is an integer at $t \sim 4.8$ s [38]. Figure 14 shows that in the same JET discharge the ITB triggering event, observed as an increase of $T_e$ in the region close to $q_{\min}(R)$ happens earlier, at $t \sim 4.6$ s. This sequence of events is characteristic of the majority of JET discharges showing that the formation of ITB just before $q_{\min} = integer$ is more likely to be associated with the depletion of rational magnetic surfaces [40], rather than with the presence of an integer $q_{min}$ value itself. Similar observations have been made on DIII-D [41]. Another important example of MHD spectroscopy are the studies of sawtooth crashes on C-MOD [43, 44] and JET [45], in which ACs (RSAEs) are observed between the sawtooth crashes. Figure 15 shows the detected RSAEs on C-MOD between two sawtooth crashes, which convincingly indicate the shear reversal inside the $q = 1$ radius. Figure 16 shows the relevant reconstruction of the $q(r)$-profile from the modes observed.

The use of MHD spectroscopy has become routine for JET, DIII-D, NSTX, MAST, and ASDEX-Upgrade, and the extension to 3D plasmas is being implemented on LHD [42].

## 5. The near-threshold nonlinear theory of frequency sweeping modes

The FS scenarios of energetic particle-driven Alfvén instabilities were commonly observed on DIII-D, JT-60U, ASDEX-Upgrade, MAST, NSTX, START, and LHD machines with NBI heating (see, e.g. [46] and References therein). In contrast to FL scenarios, neither frequency nor structure of FS modes is determined by the bulk plasma equilibrium during the nonlinear mode evolution. Description of FS modes is essentially nonlinear, and the linear MHD spectral codes have very limited applicability.

The recent progress in describing FS instabilities is associated with kinetic theory [47] of energetic particle-driven waves with different collisional effects [48], drag and diffusion, replenishing the unstable distribution function and satisfying the near-threshold condition

$$\gamma \equiv \gamma_L - \gamma_d << \gamma_d \leq \gamma_L. \tag{8}$$

In this limit, a lowest order cubic nonlinear equation for the mode amplitude describes "soft" nonlinear FL scenarios (steady-state, pitchfork splitting, and chaotic) when diffusion dominates at the wave-particle resonance in the phase space, and "hard" (explosive) nonlinear scenario when the drag dominates or the diffusion characteristic time is much longer than $\gamma^{-1}$. The explosive mode evolution goes beyond the cubic nonlinearity and the fully nonlinear model shows a spontaneous formation of long-living structures, holes and clumps, in the energetic particle distribution [49]. These structures are nonlinear energetic particle modes, which travel through the phase space and sweep in frequency [50] exhibiting many of the characteristics of FS modes seen in experiments [46].

Among the variety of the frequency sweeping spectra obtained in modelling [51,52], the long range frequency sweeping phenomenon attracts most attention, due to its relevance to the experimental observations. Figure 17 shows results of MAST experiment with super-Alfvénic NBI driving Alfvén instability when the resonance $V_{\|beam} = V_A$ is in phase space region dominated by electron drag of the beam ions. It is seen, that similarly to Figure 2, some of the modes sweep in frequency to a very long range of $|\delta\omega/\omega| \cong 0.5$. Although modelling with HAGIS code [30] reproduces the characteristic spectrum observed in experiments as Figure 18 shows, the range of the frequency sweeping is not as large as that observed on MAST. New theoretical approaches are being developed for long-range FS modes [50, 53, 54].

The dominant transport mechanism for nonlinear FS modes is convection of particles trapped in the wave field. This mechanism is also characteristic for strongly unstable energetic particle modes that are already non-perturbative in the linear regime [55]. Experimentally, validation of the hole-clump formation and transport was made with an NPA diagnostic on LHD [56]. Figure 19 shows how the flux of energetic beam ions sweeps in energy together with FS modes. A new theory of continuous hole-clump triggering [57] shows that a single resonance can produce transport higher than the quasi-linear estimate, due to the convection of the resonant ions trapped in the field of a travelling wave.

A joint ITPA experiment validating the near-threshold model is in progress, with MAST and LHD comparison indicating that the parameter space for bursting AEs shrinks for core-localised global AE (GAE) on LHD, in which GAEs exist because of a $q(r)$-profile different from that in tokamak [42]. In parallel, study of experimental data continues. On NSTX bursting FS TAEs were observed in the form of "avalanches" consisting of several coupled modes with strong downward frequency sweep and amplitudes higher than un-coupled TAEs [58]. The experimentally observed ~10% drops in the neutron rate during the avalanches were explained by a decrease in the beam energy and losses resulting from interaction with TAEs.

## 6. Possible control of Alfvén instabilities in burning plasmas using ECRH

The problem of controlling Alfvén instabilities and fast ion transport caused by AEs is one of the important avenues for future exploitation in both experiment and theory. The most encouraging results in this area were obtained on DIII-D, where ECRH was found to suppress RSAEs excited by the beam ions [59]. A direct comparison of ECCD effect versus ECRH [59] has shown that it is the heating, not the current drive, which provides the mode suppression, possibly via electron pressure gradient or via increased damping due to larger population of trapped electrons. A new joint ITPA experiment was set up in order to assess such effect on DIII-D, ASDEX-Upgrade, TCV, LHD, TJ-II, HL-2A, and KSTAR. From the standpoint of targeting and affecting a particular type of waves with a known location, ECRH is an ideal tool since it can provide highly localised targeted power deposition on ITER. Figure 20 shows the interferometry data on RSAE activity in DIII-D discharges with ECRH. The amplitudes and number of unstable AEs decreases when ECRH is applied to the localisation region of RSAEs at $q_{min}$.

In ITER with possible high-$n$ TAEs occupying a wide radius, ECRH, due to its high localisation, may suppress TAEs in a narrow region rather than in whole plasma. However, if the width of the TAE-free zone is larger than the orbit width of the energetic ions, this zone could become a transport barrier for the TAE-induced transport of the energetic ions. With the expertise gained in ECRH triggering of ITBs for thermal plasma, the possibility of employing ECRH for creating TAE-free transport barriers for energetic particles in ITER could be feasible. Further study of ECRH suppression of AEs is required.

## 7. Conclusions

In summary, a systematic and significant recent effort in diagnosing the energetic ion driven Alfvén instabilities and related transport of the energetic ions in JET, DIII-D, ASDEX Upgrade (AUG), Alcator C-MOD, LHD, NSTX and MAST is setting the stage for a new understanding of such instabilities relevant to the next-step burning plasma experiment. Based on the detailed measurements of AEs on several machines, their amplitude and mode structure, successful modelling was performed reproducing both qualitative and quantitative effects of the energetic particle transport. The observed discrete spectra of AEs with frequencies locked to the plasma equilibrium provide a reliable MHD spectroscopy tool for providing information on plasma equilibrium on many machines thus demonstrating the correctness of the modelling codes under various conditions. For the frequency sweeping instabilities the role of collisional effects during the nonlinear mode evolution provides a credible explanation of the main experimentally observed features although a theory-to-experiment comparison in realistic geometry with accurate collisional operator has yet to be completed.

These results obtained on different machines world-wide display both the diagnostic and modelling requirements for the next-step burning plasma experiment as well as the importance of theoretical development of new essential phenomena of the burning plasma.

**Acknowledgments**

This work, supported by the European Communities under the contract of Association between EURATOM and CCFE, was carried out within the framework of the European Fusion Development Agreement. The views and opinions expressed herein do not necessarily reflect those of the European Commission. This work was also part-funded by the RCUK Energy Programme under grant EP/I501045.

**Figure captions**

**Fig.1** Magnetic spectrogram of ICRH-driven Alfvén instabilities in JET discharge #40329 with gradually increasing ICRH power.

**Fig.2** (a) Magnetic spectrogram of NBI-driven Alfvén instabilities in JT-60U discharge E36379 (B=1.2 T, $E_{NBI}$=360 keV); b) Mirnov coil signal.

**Fig.3** DIII-D discharge 122117, t = 410.6 ms, on axis: $T_e$=1.5 keV, $T_i$= 1.9 keV, $B_T$ = 2.0 T, $n_e$ =2.17x $10^{13}$ cm$^{-3}$, $f_{rot}$=9.9 kHz. (a) Radial profile of ECE power spectra. RSAEs (blue line) and TAEs (red) are pointed out. The solid overlaid line is the q(r)-profile and the dashed line is $T_e$(r) profile. (b) n = 3 Alfvén continuum including toroidal rotation. The horizontal lines mark n= 3 RSAE (blue) and TAE (red).

**Fig.4** Diamonds show ECE radiometer measured temperature perturbations for RSAE (left) and TAE (right) from Figure 3; solid line − the NOVA predictions. The eigenmode amplitude is obtained by least squares fit to the ECE data.

**Fig.5** Top: Power waveforms of ICRH and NBI in JET discharge #74951. Middle: Temporal evolution of $T_e$ measured with multichannel ECE. Central channel corresponds to $T_e$ at R = 3.03m, the sawtooth inversion radius is at R = 3.35m. Bottom: time evolution of the DD neutron yield.

**Fig.6** Tornado modes detected with Mirnov coils (top) and FIR interferometry (bottom) in JET discharge # 74951.

**Fig.7** Lines-of-sight of the 2D gamma-camera on JET.

**Fig.8** Time evolution of the gamma-ray signals for channels 14 - 18 in JET discharge # 74951. The signals in central channels (15, 16) decrease, while the signals in outer channels (14,18) increase, showing the redistribution of gamma-rays from energetic deuterons during Alfvén instability.

**Fig.9** Gamma intensity in the 19 channels of gamma-camera (JET pulse # 74951). Here we show measured pre-TAE (blue) and during TAE (green) profiles. Simulated gamma intensity is shown in red (initial data) and black (after redistribution).

**Fig.10** Fast-ion pressure profiles and FIDA density profiles at two different times that correspond to normalized neutron rates of 0.66 and 0.94. The dashed lines are the classical pressure profile predicted by TRANSP. The FIDA density profile is normalized to the MSE-EFIT $p_f$ profile at 1.2 s.

**Fig.11** Radial profiles of the beam ions from the ORBIT code with AEs versus the experimental data.

**Fig.12** The beam ion radial profiles shown in pairs: the TRANSP profiles (without AE) versus the profiles from the critical radial gradient approach as determined by AE instability in the quasi-linear model.

**Fig.13** ACs detected with interferometry in JET discharge #61347. Grand AC is seen at t ~ 4.8 s.

**Fig.14** $T_e(t)$ measured with multi-channel ECE in JET discharge #61347. An ITB triggering event is seen at t ~ 4.6 s.

**Fig.15** PCI spectrogram showing the modes following the second sawtooth crash in C-MOD. NOVA calculation of the frequency spectra are overlaid in dashed line and was used to infer evolution of $q_{min}$ from 0.99–0.92 over the period of RSAE.

**Fig.16** Left: The n= 3 Alfvén continuum calculated for the q profile shown at bottom right, with $q_0$=0.96 and $q_{min}$=0.95 at r/a=0.25 corresponding to the conditions near t=0.275 s in Fig. 15. Top right: An n= 3 RSAE computed with NOVA. The q profile is shown on the bottom right for reference.

**Fig.17** Spectrogram showing FS Alfvén modes driven by NBI in MAST discharge #27177.

**Fig.18** Nonlinear HAGIS simulation of Alfvén instability on MAST driven by slowing-down beam distribution function.

**Fig.19** Top: Mirnov coil data; Middle: magnetic spectrogram; Bottom: Time evolution of tangential energetic neutral spectrum. The NPA viewing angle is set to $0^0$.

**Fig.20** Windowed crosspower spectra of vertical and radial $CO_2$ interferometer data for 1.9MW ECRH deposition at (a) near axis in #128564, (b) near $q_{min}$ in #128560. Overlayed white curves are a typical RSAE and the local TAE frequency.

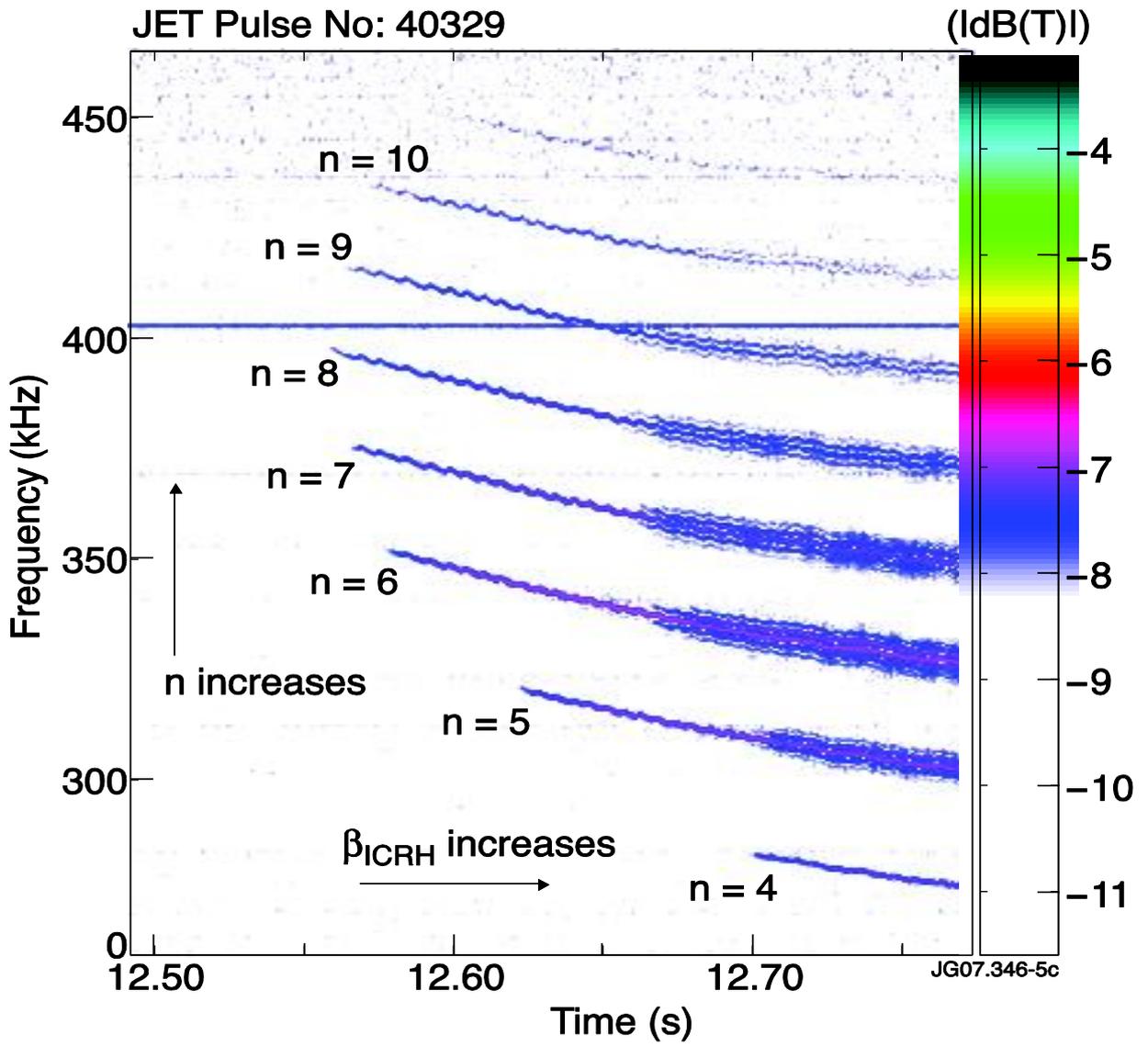

Figure 1

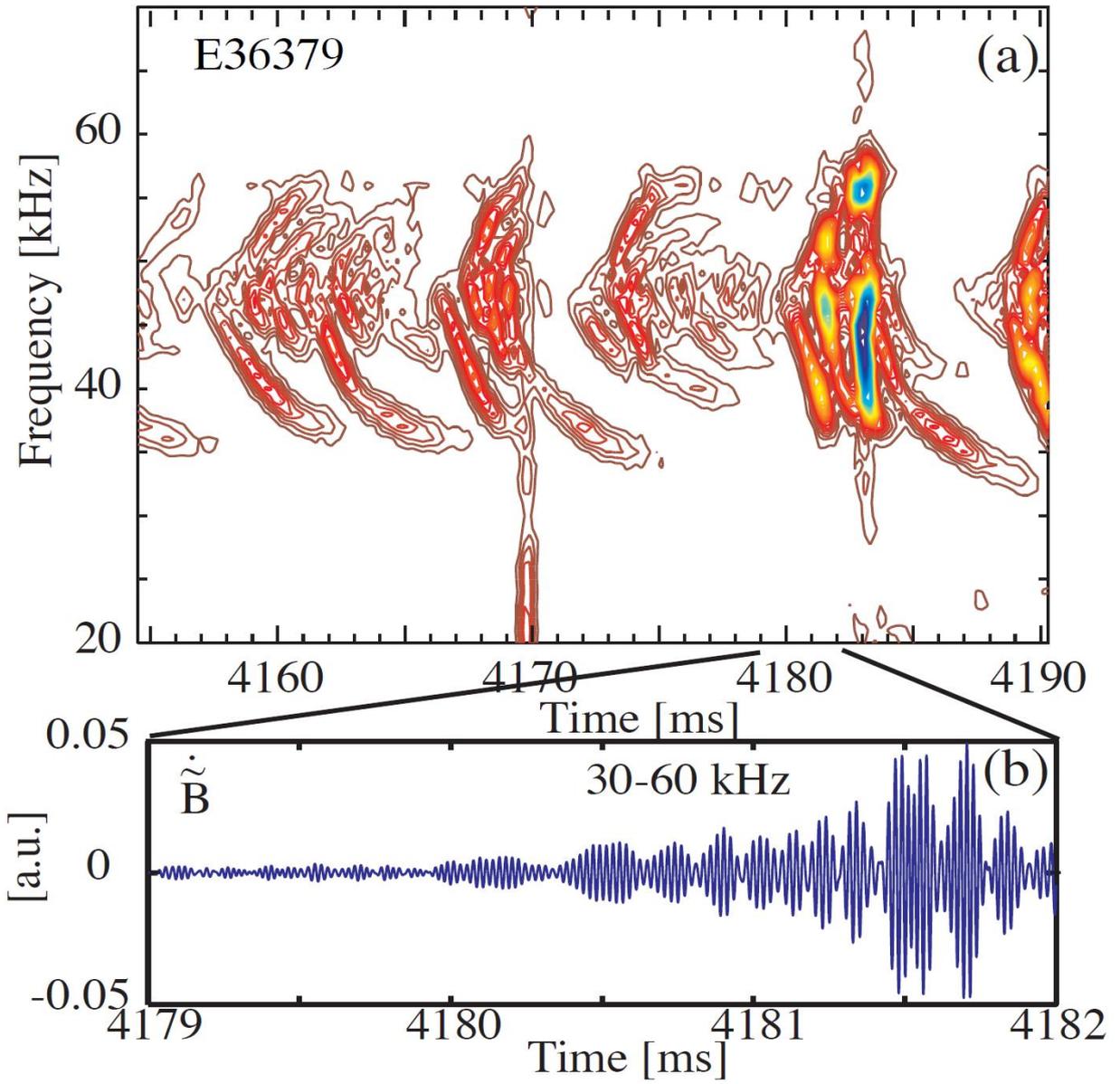

Figure 2

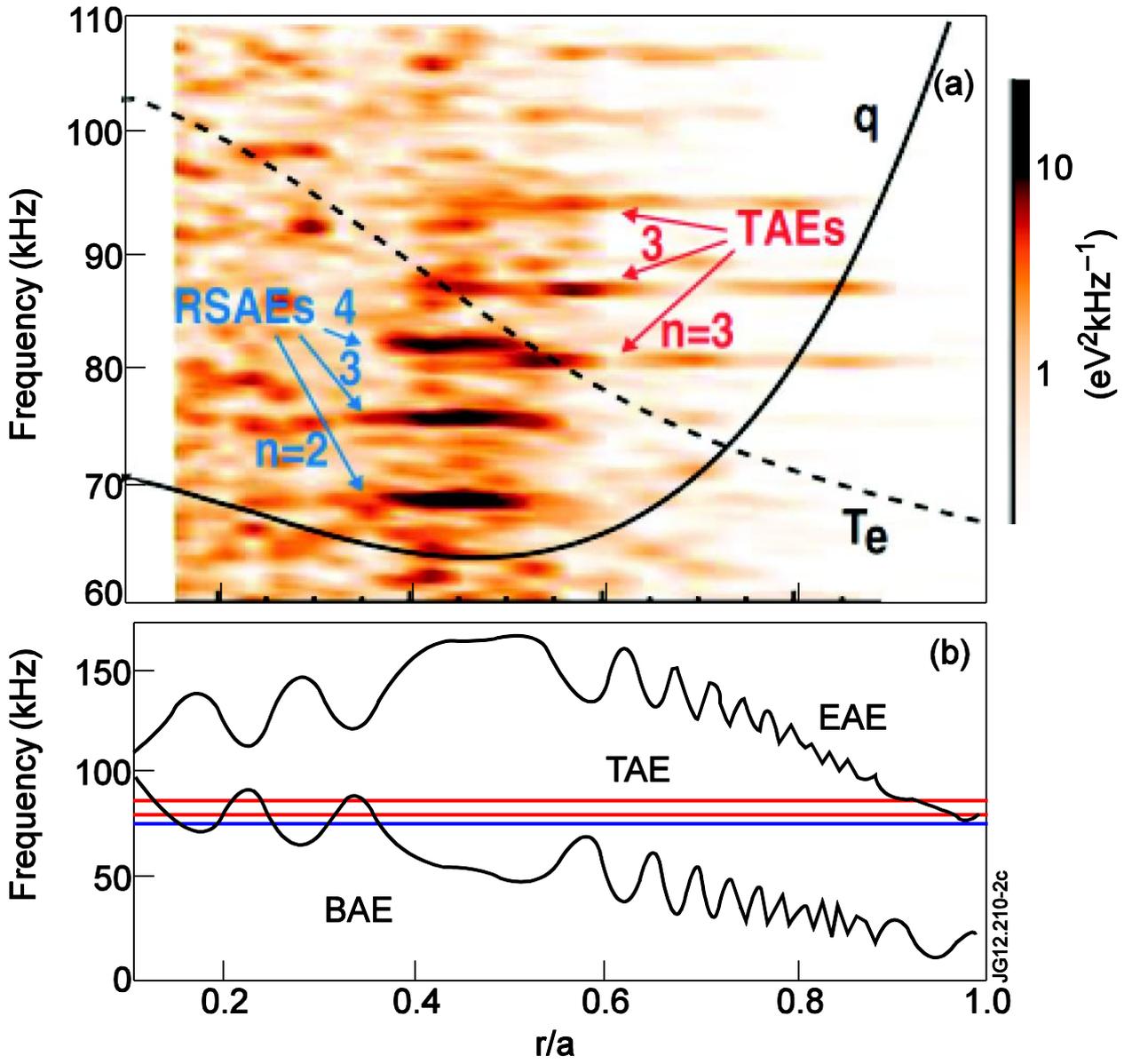

Figure 3

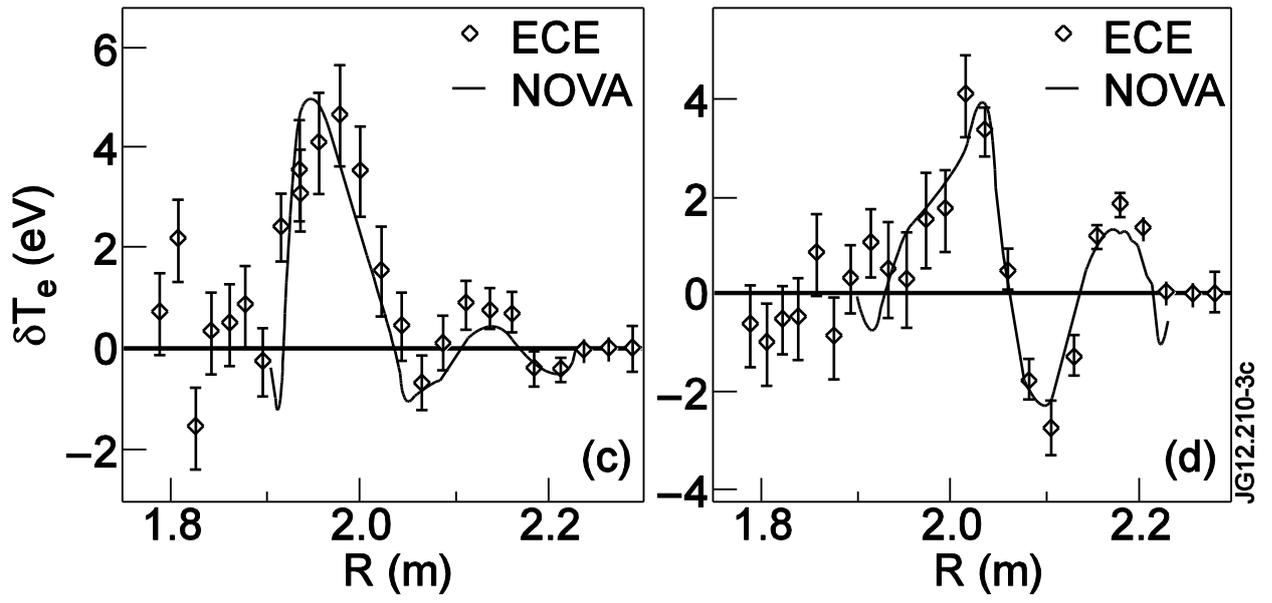

Figure 4

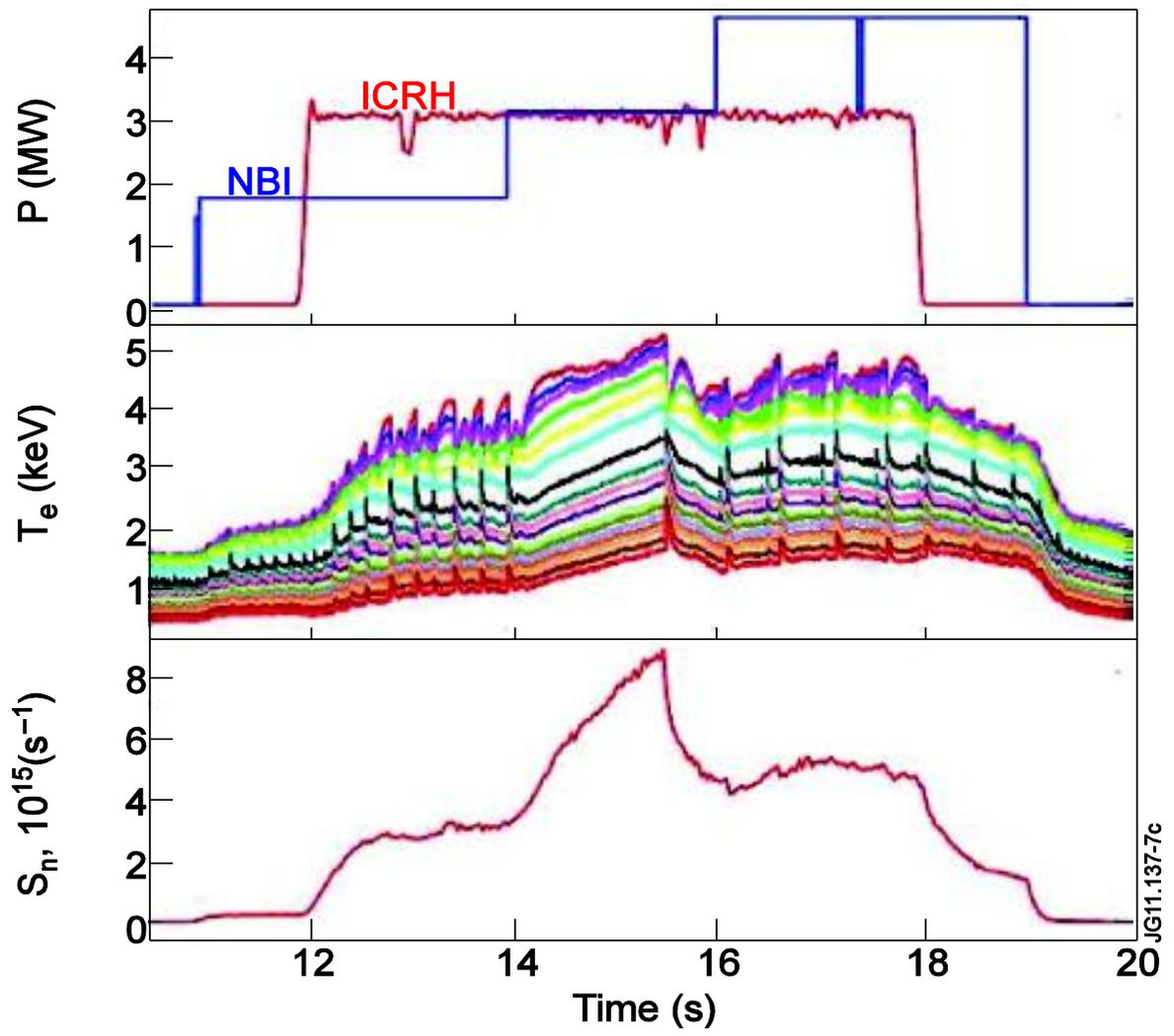

Figure 5

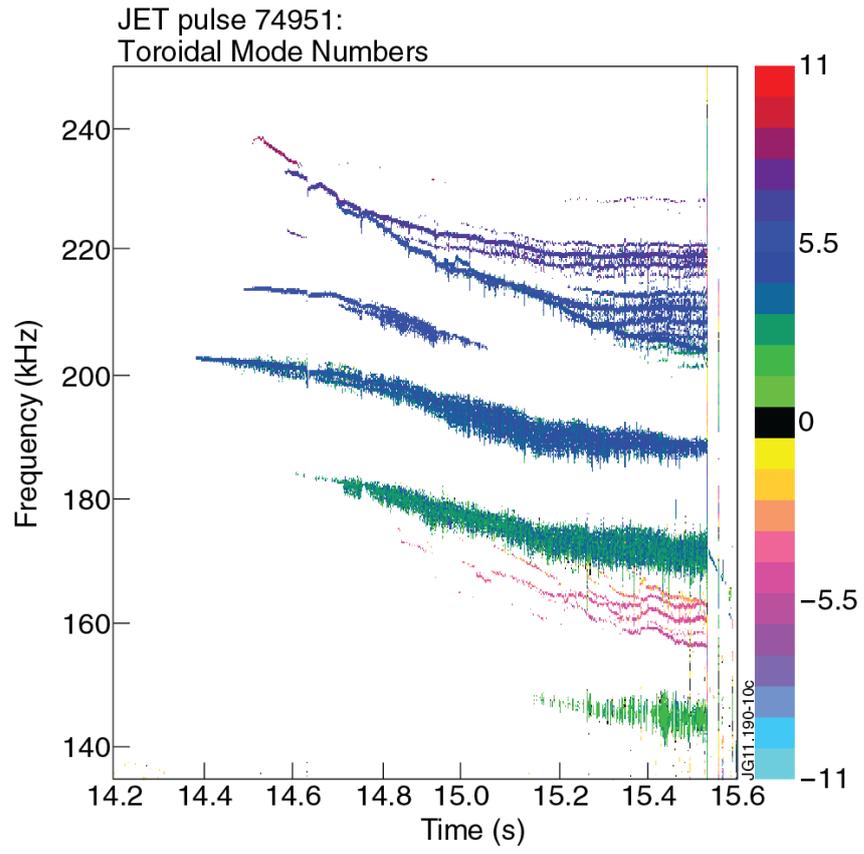

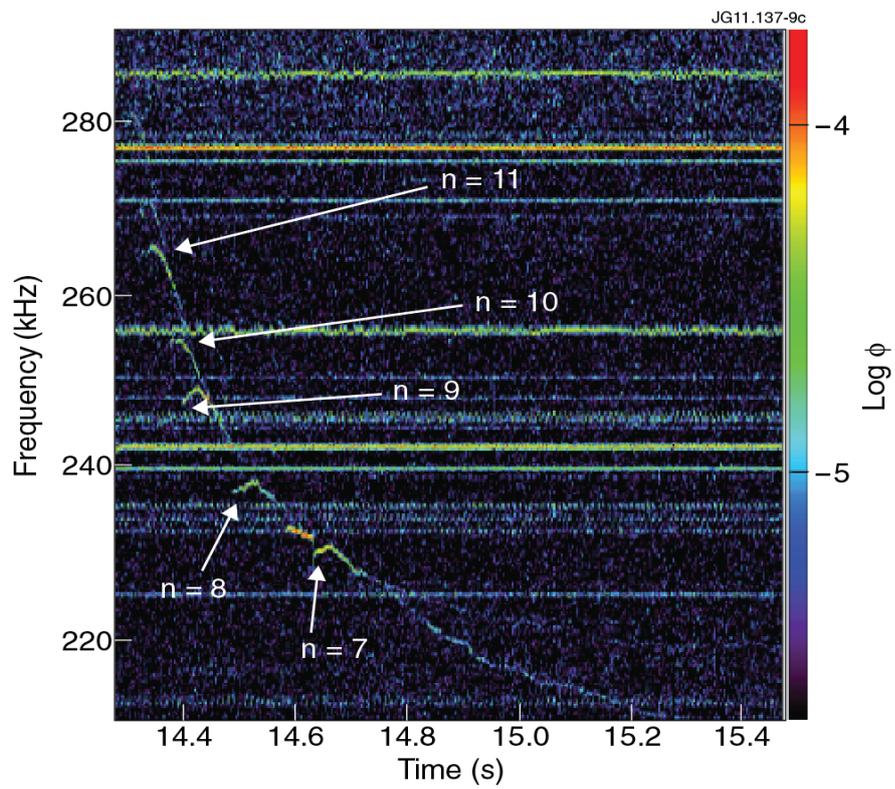

Figure 6

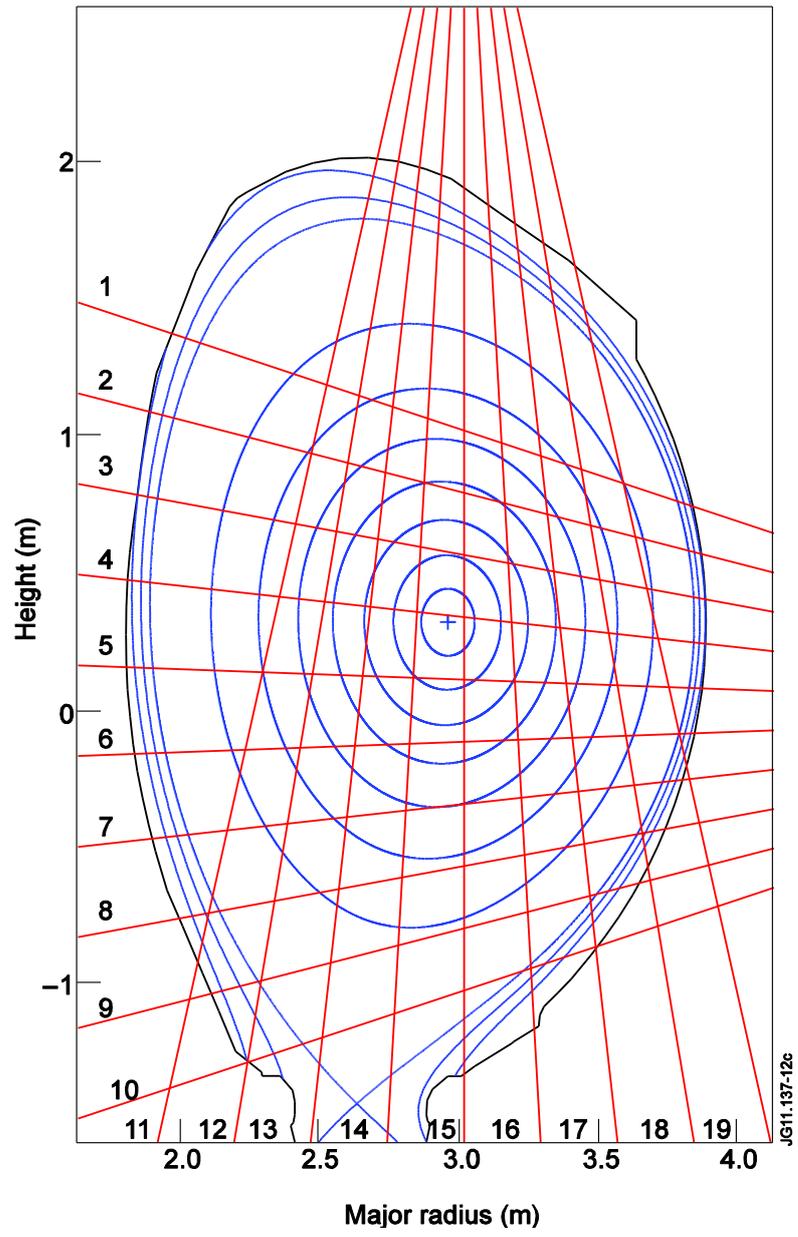

Figure 7

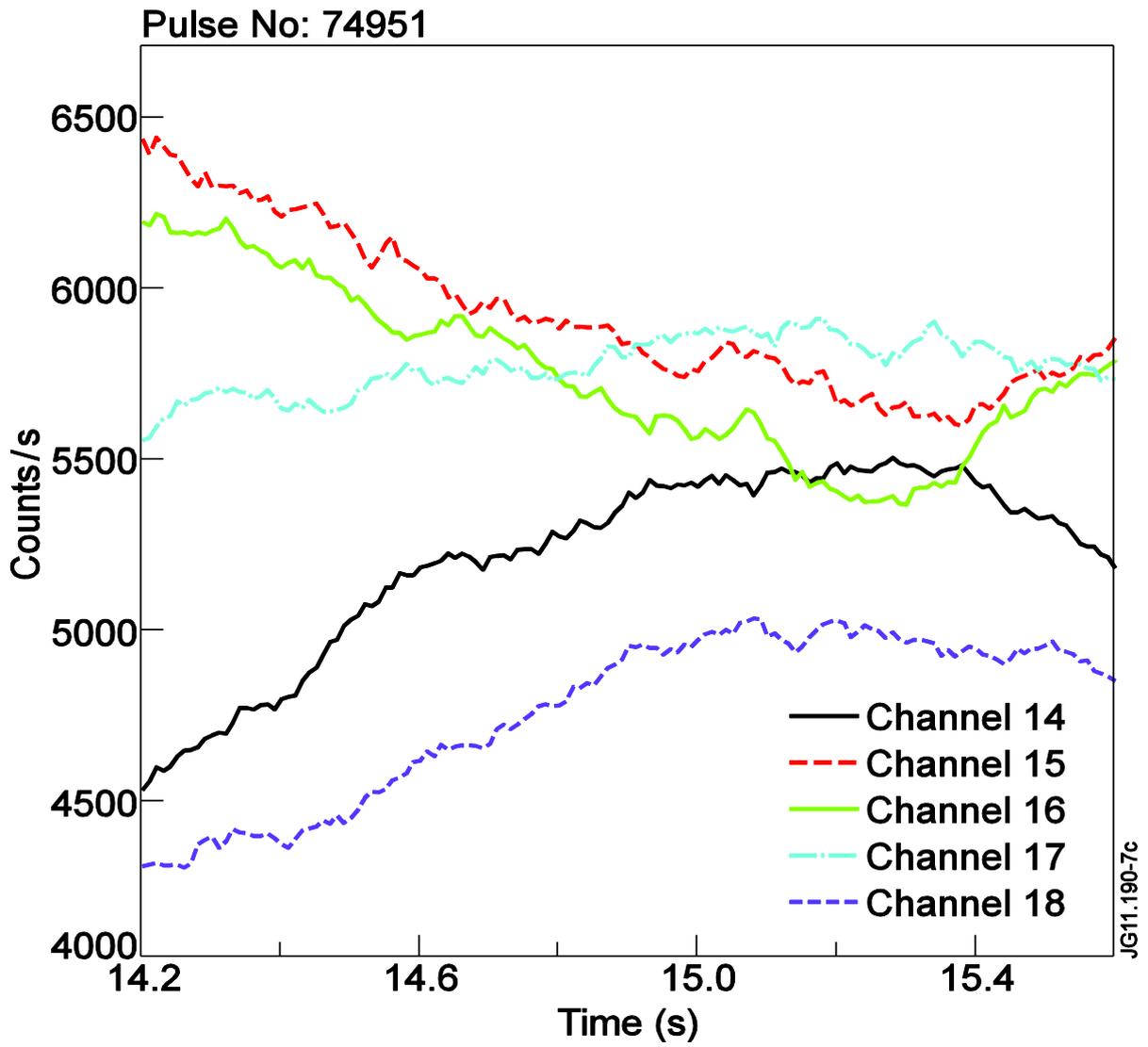

Figure 8

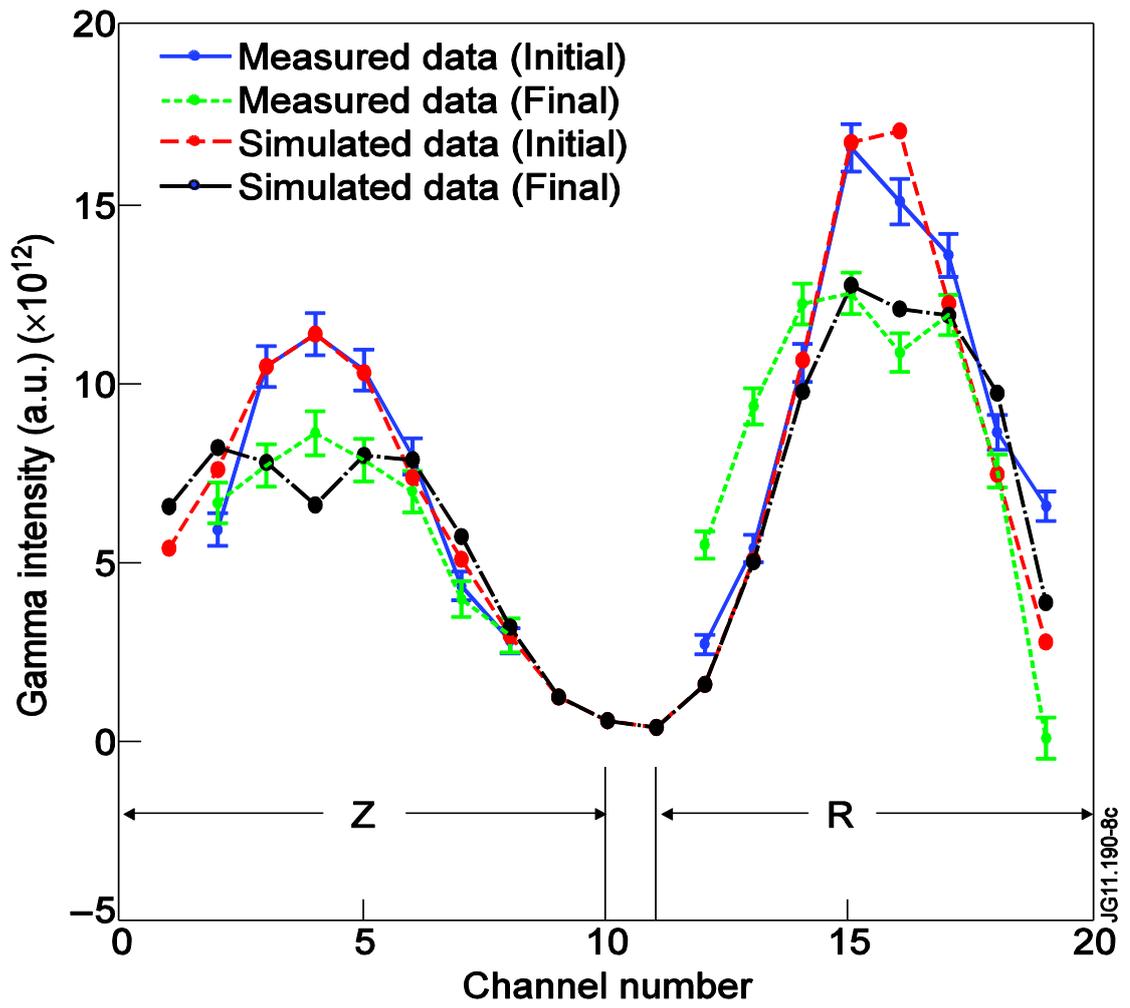

Figure 9

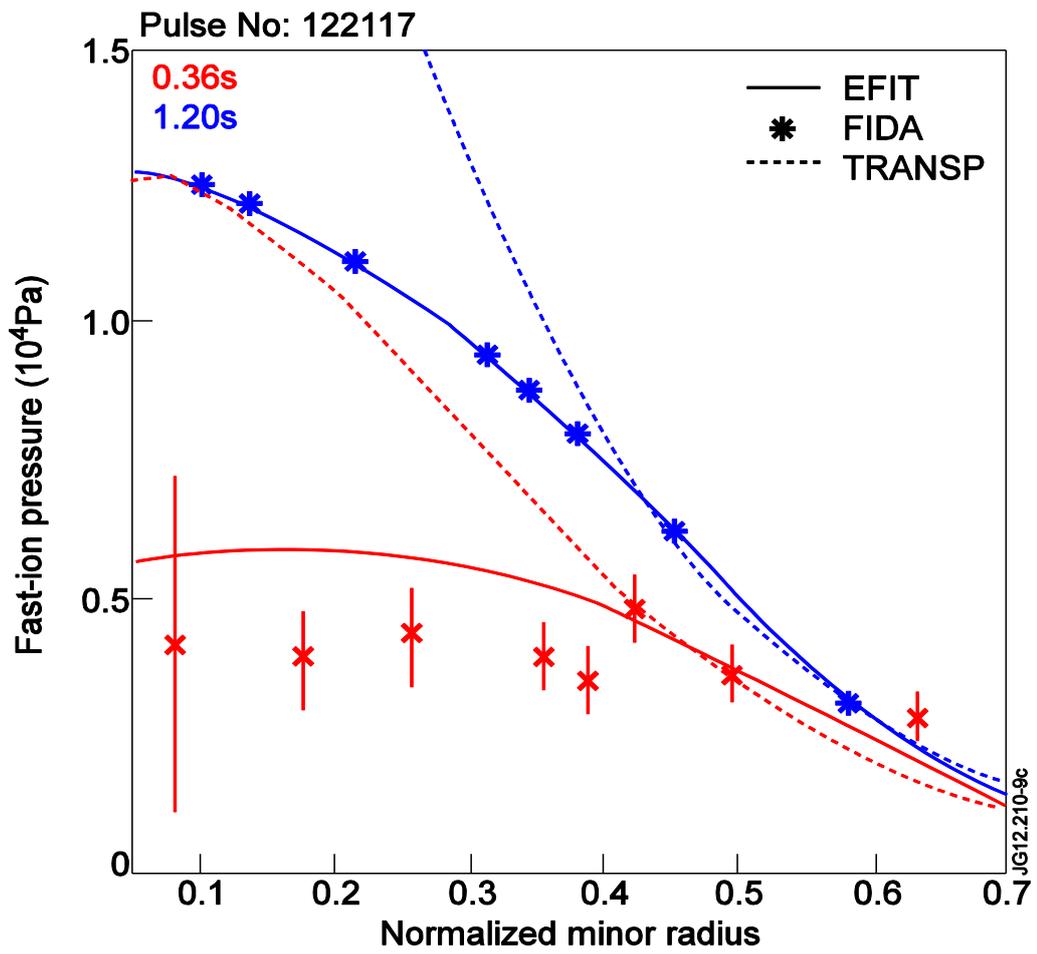

Figure 10

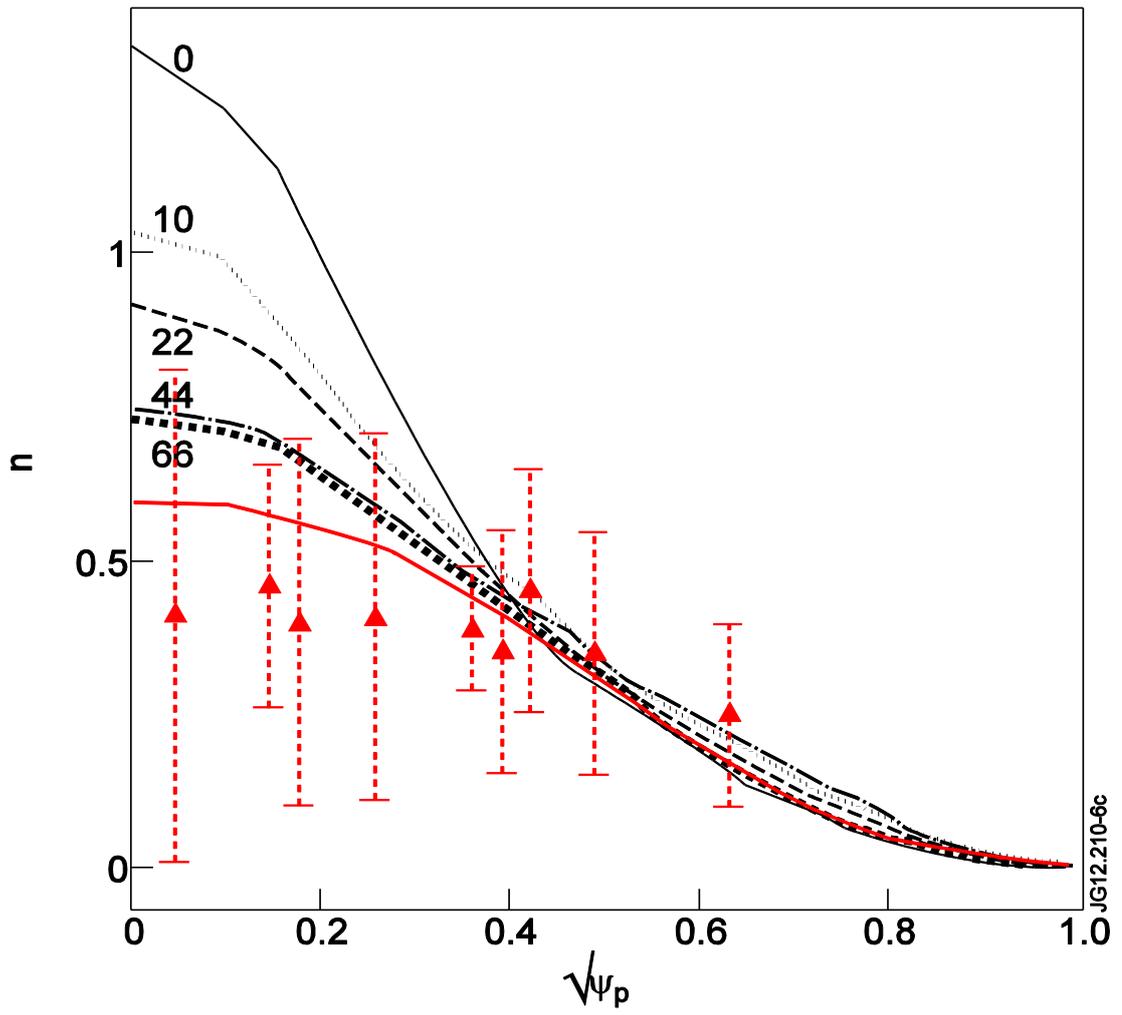

Figure 11

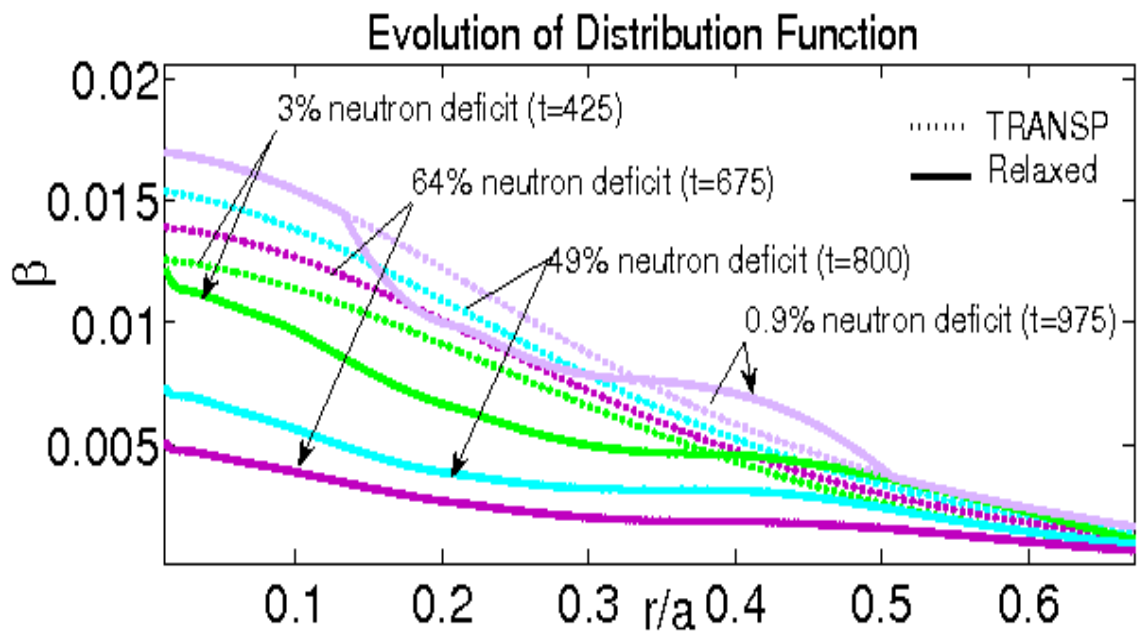

Figure 12

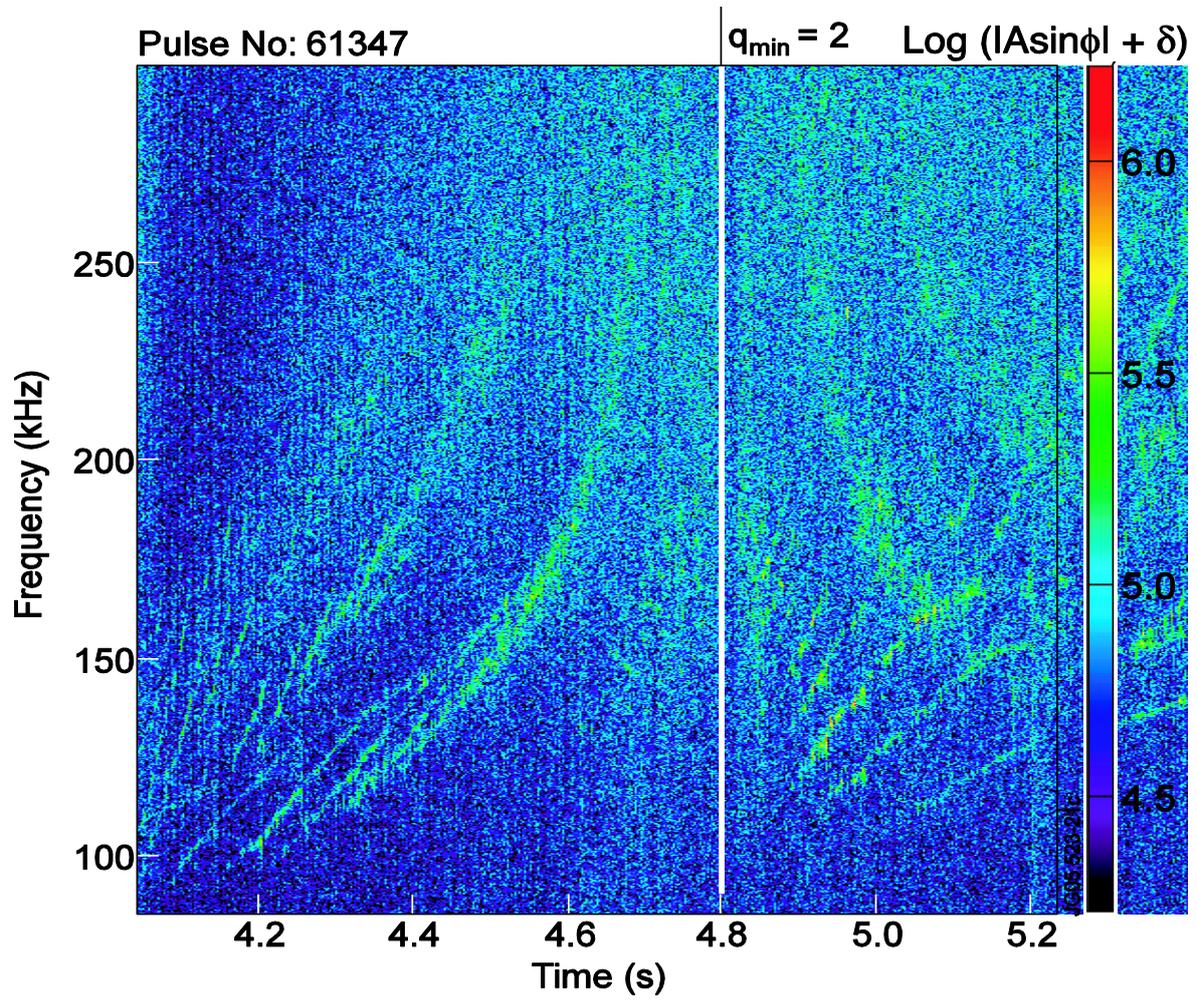

Figure 13

Figure 14

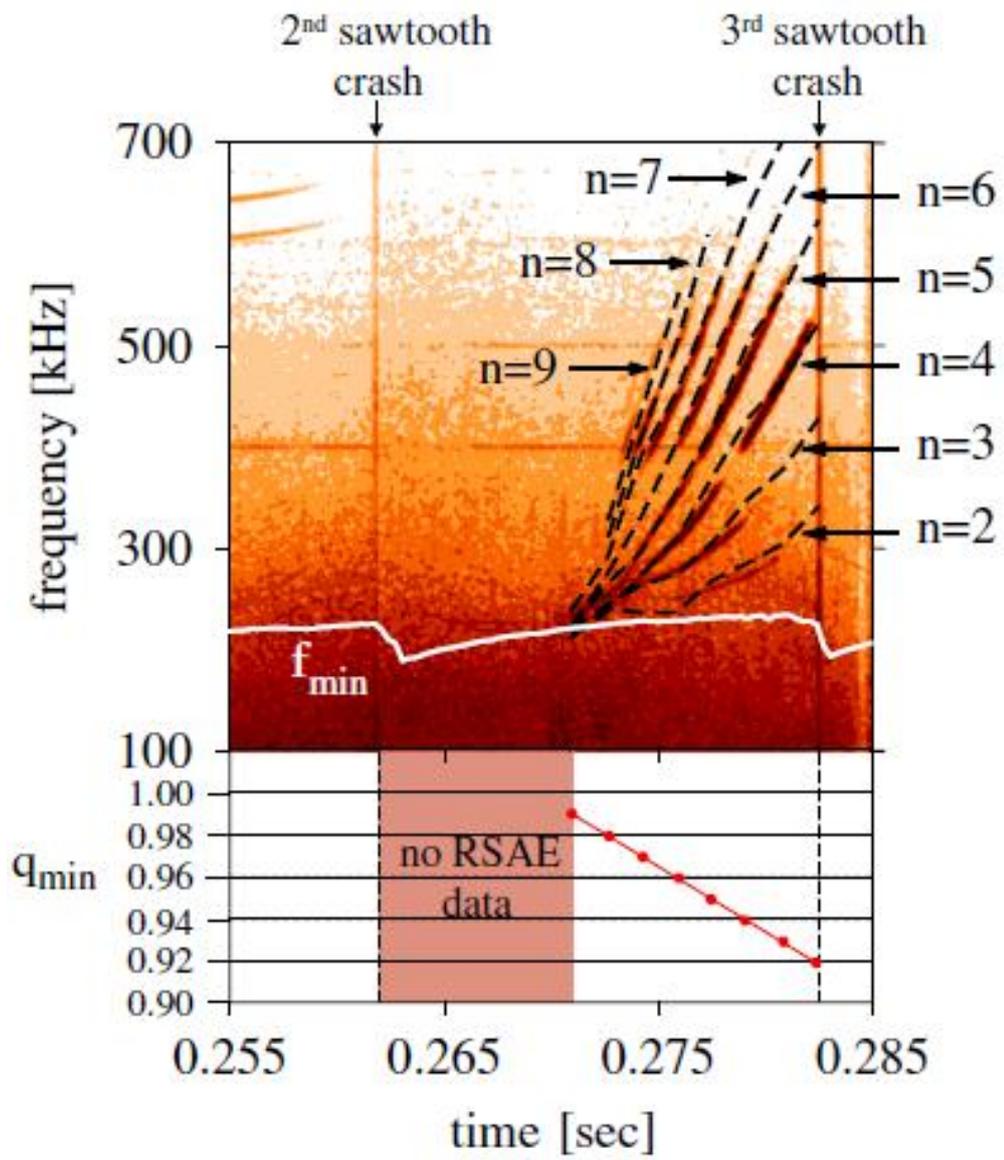

Figure 15

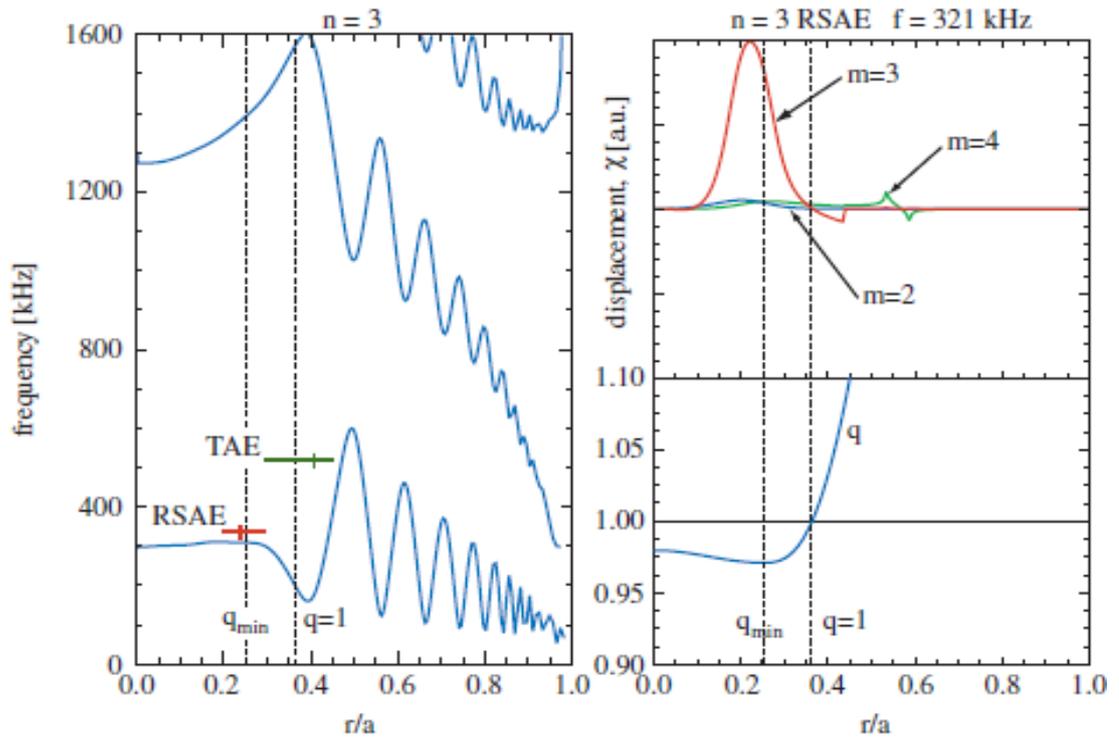

Figure 16

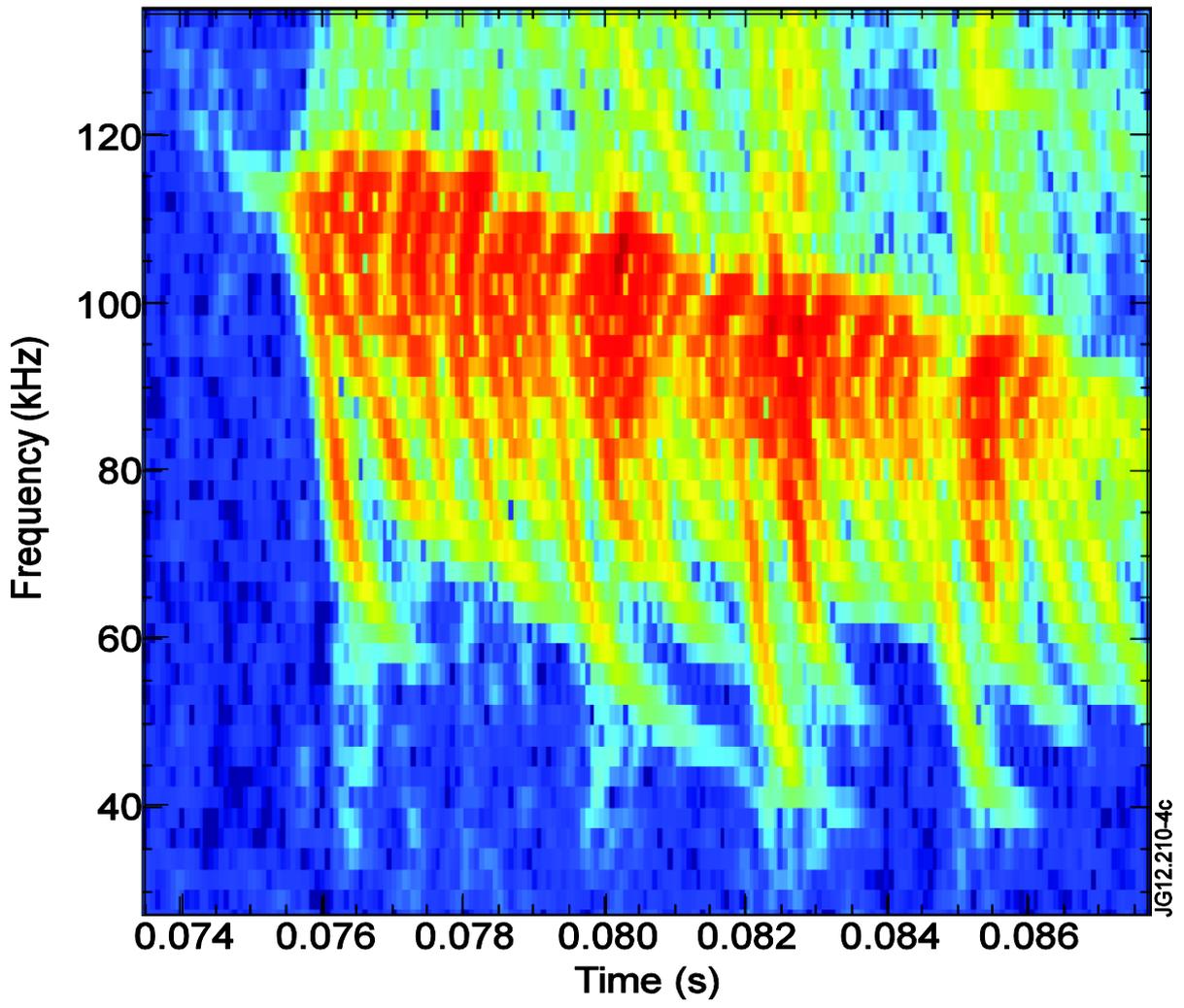

Figure 17

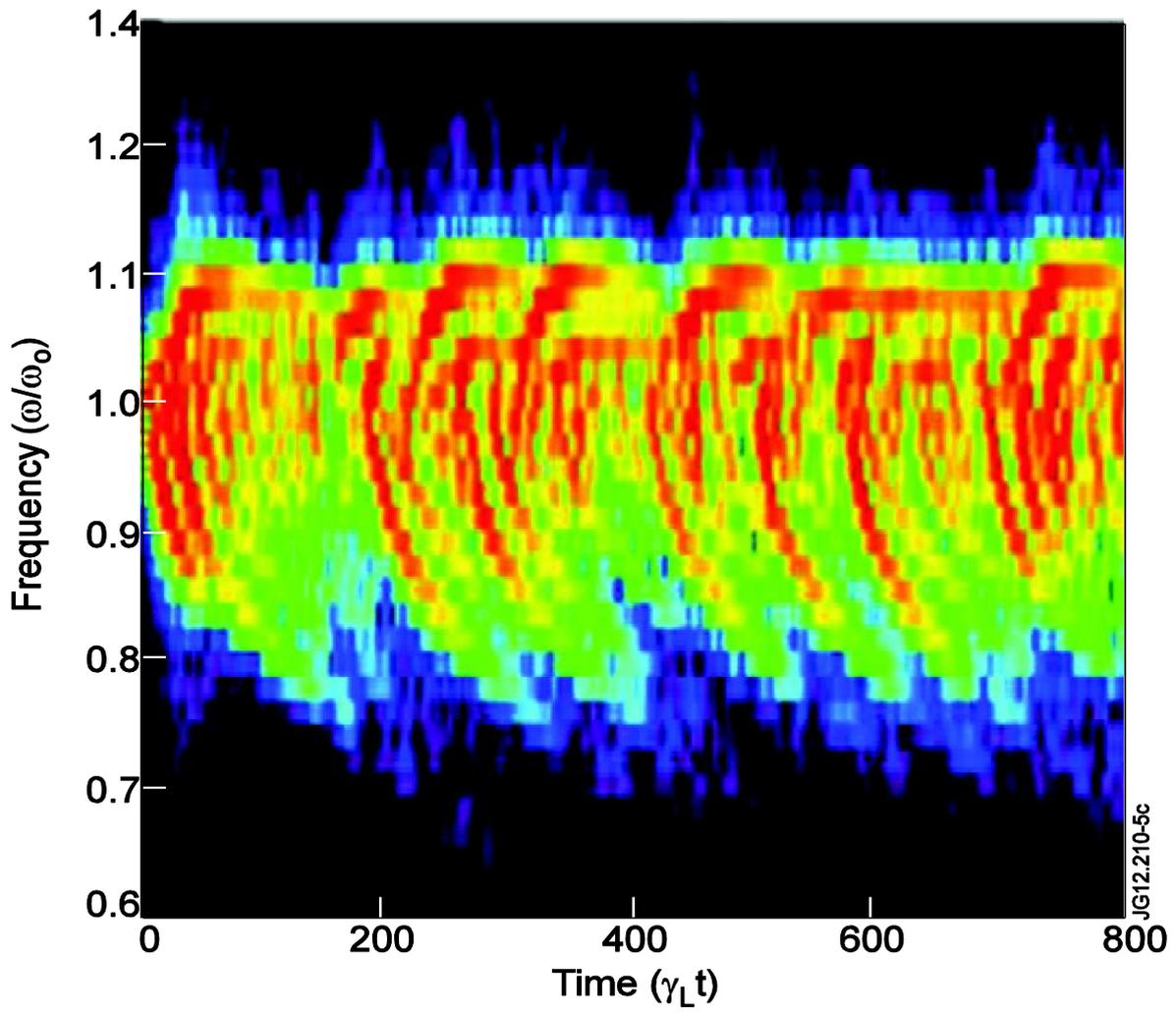

Figure 18

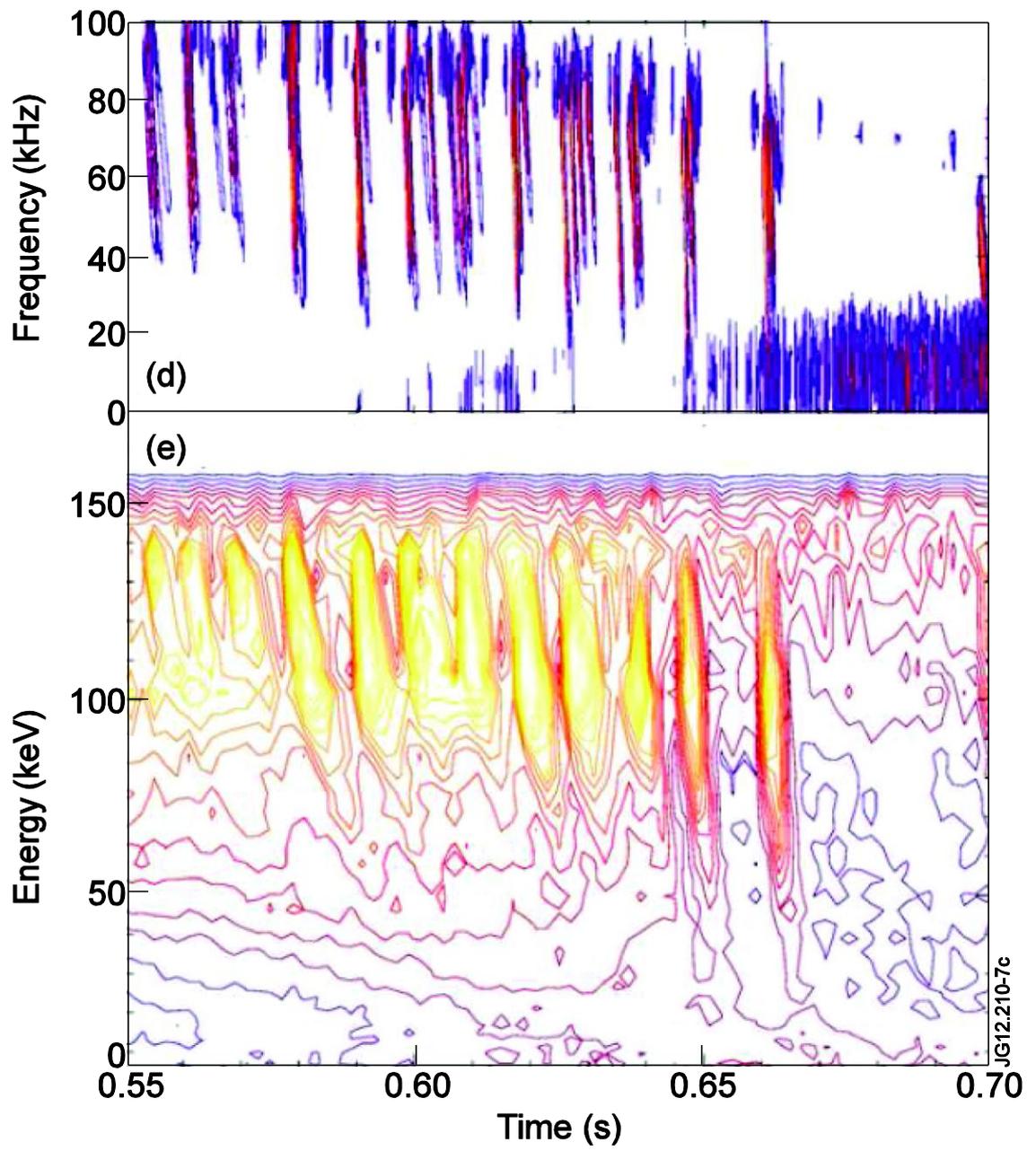

Figure 19

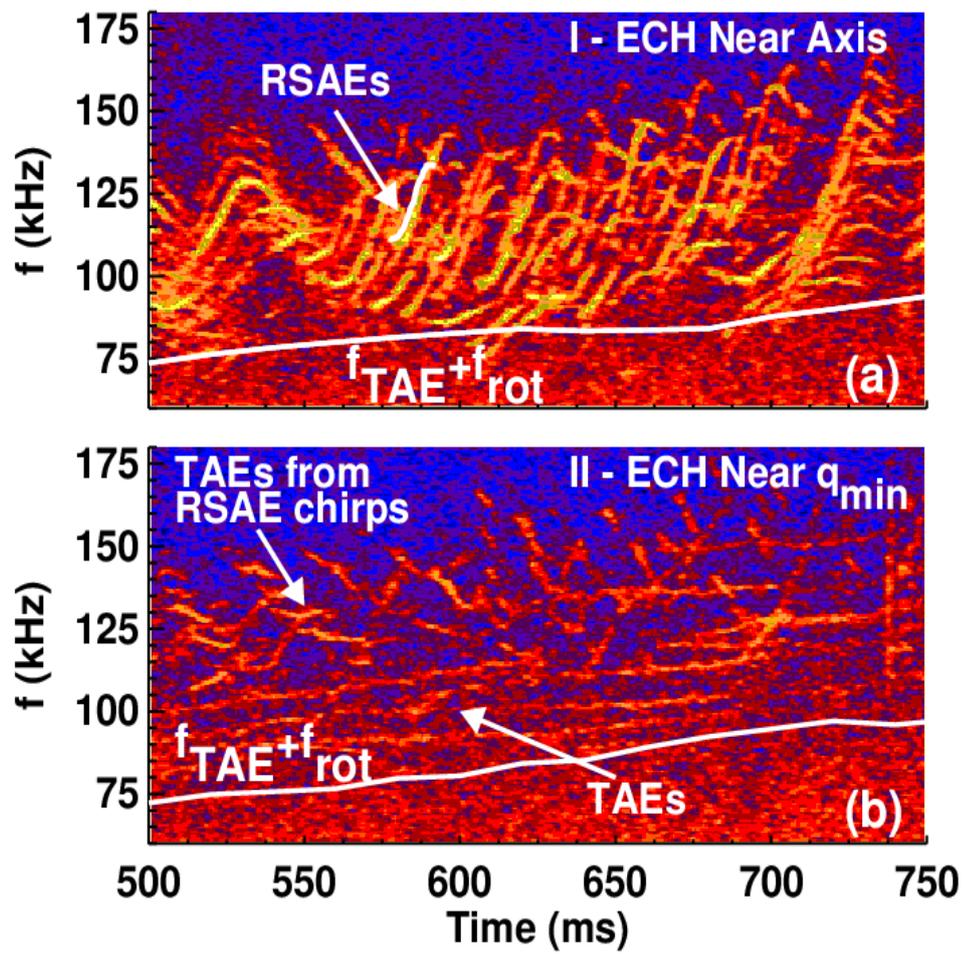

Figure 20